\documentclass[reprint,aps,pra,showpacs,floatfix,onecolumn]{revtex4-1}
\usepackage{graphicx}
\usepackage{dsfont}
\usepackage{epstopdf}

\newcommand{\be}{\begin{equation}}
\newcommand{\ee}{\end{equation}}

\begin{document}

\title{Limit density of 2D quantum walk: zeroes of the weight function}

\author{M. \v Stefa\v n\'ak\email[correspondence to:]{martin.stefanak@fjfi.cvut.cz}}
\affiliation{Department of Physics, Faculty of Nuclear Sciences and Physical Engineering, Czech Technical University in Prague, B\v
rehov\'a 7, 115 19 Praha 1 - Star\'e M\v{e}sto, Czech Republic}

\author{I. Bezd\v ekov\'a}
\affiliation{Department of Physics, Faculty of Nuclear Sciences and Physical Engineering, Czech Technical University in Prague, B\v
rehov\'a 7, 115 19 Praha 1 - Star\'e M\v{e}sto, Czech Republic}

\author{I. Jex}
\affiliation{Department of Physics, Faculty of Nuclear Sciences and Physical Engineering, Czech Technical University in Prague, B\v
rehov\'a 7, 115 19 Praha 1 - Star\'e M\v{e}sto, Czech Republic}

\begin{abstract}
Properties of the probability distribution generated by a discrete-time quantum walk, such as the number of peaks it contains, depend strongly on the choice of the initial condition. In the present paper we discuss from this point of view the model of the two-dimensional quantum walk analyzed in K. Watabe et al., Phys. Rev. A {\bf 77}, 062331, (2008). We show that the limit density can be altered in such a way that it vanishes on the boundary or some line. Using this result one can suppress certain peaks in the probability distribution. The analysis is simplified considerably by choosing a more suitable basis of the coin space, namely the one formed by the eigenvectors of the coin operator.
\end{abstract}

\maketitle

\section{Introduction}
\label{Introduction}

Quantum walks \cite{adz,meyer,fg} were proposed as extensions of the concept of a classical random walk to the unitary evolution of a quantum particle on a discrete graph or lattice. They have found  promising applications in quantum information processing, e.g. in search algorithms \cite{skw}, graph isomorphism testing \cite{gamble}, finding
structural anomalies in graphs \cite{cottrell}, and perfect state transfer \cite{kendon:qw:pst}. Moreover, quantum walks were shown to be universal tools for quantum computation \cite{childs}.

Suitable tools for the analysis of homogeneous quantum walks on infinite lattice are the Fourier transformation \cite{ambainis} and the weak-limit theorems \cite{Grimmett}. While the properties of many quantum walks on a line are well understood \cite{Konno:2005,konno:wigner,falkner,stef:limit}, less is know about quantum walks on higher-dimensional latices. Indeed, there are many technical difficulties, e.g. diagonalization of the evolution operator. One of the few models of 2D quantum walks which is well understood is the one analyzed in \cite{watabe:grover}. This model is a one-parameter extension of the 2D Grover walk which preserves its key feature, namely the trapping effect (or localization) \cite{inui}. The coin parameter controls the area covered by the quantum walk, which in general is an elliptic disc and reduces to a circle for the 2D Grover walk.

In the present paper we focus on the role of the initial conditions on the shape of the probability distribution resulting from the 2D quantum walk of \cite{watabe:grover}. We are interested in initial states which lead to non-generic probability distributions, such as those with reduced number of peaks. In order to find them we first simplify the results of \cite{watabe:grover} by converting them to a more suitable basis of the coin space. Following \cite{stef:limit} we choose the basis formed by the eigenvectors of the coin operator. We then discuss various initial coin states which result in non-generic probability distribution. In particular, we show that the limit density can be set to zero on some line. This can be used to suppress peaks in the probability distribution.

The paper is organized as follows: First, in Section~\ref{sec2} the results of \cite{watabe:grover} are briefly reviewed. Next, we convert them into more suitable basis to simplify the following analysis. In Section~\ref{sec3} various initial states which lead to non-generic probability distributions are discussed. We conclude and present an outlook in Section~\ref{sec4}.

\section{2D quantum walk}
\label{sec2}

Let us first briefly review the results of \cite{watabe:grover}. The authors have considered a quantum walk on a two-dimensional square lattice where the particle can in each step move from its present position $(x,y)$ to the nearest neighbours $(x\pm 1,y)$ and $(x,y\pm 1)$. These displacements correspond to the four states $|R\rangle$ , $|L\rangle$, $|U\rangle$ and $|D\rangle$ which form the standard basis of the coin space ${\cal H}_C$. In this standard basis the coin operator is given by the following matrix
\begin{equation}
\label{coin}
C = \left(
\begin{array}{cccc}
 -p & 1-p & \sqrt{p(1-p)} & \sqrt{p(1-p)} \\
 1-p & -p & \sqrt{p(1-p)} & \sqrt{p(1-p)} \\
 \sqrt{p(1-p)} & \sqrt{p(1-p)} & p-1 & p \\
 \sqrt{p(1-p)} & \sqrt{p(1-p)} & p & p-1 \\
\end{array}
\right),
\end{equation}
where the parameter $p$ ranges from 0 to 1. For $p=\frac{1}{2}$ the coin operator (\ref{coin}) reduces to the familiar $4\times 4$ Grover matrix. This particular model was analyzed in detail in \cite{inui}. Using the Fourier analysis and the weak limit theorem \cite{Grimmett} the authors have derived the limit density $\nu(v_x,v_y)$ of the 2D quantum walk. This allows one to evaluate the asymptotic values of all moments of re-scaled position (or pseudo-velocity) through the formula
$$
\lim\limits_{t\rightarrow +\infty}\left\langle\left(\frac{x}{t}\right)^m\left(\frac{y}{t}\right)^n\right\rangle = \int v_x^m v_y^n \nu(v_x,v_y) dv_x dv_y.
$$
The limit density of the 2D quantum walk is given by \cite{watabe:grover}
\begin{equation}
\label{lim:dens}
\nu(v_x,v_y) = \mu(v_x,v_y) {\cal M}(v_x,v_y) + \Delta \delta_0(v_x)\delta_0(v_y).
\end{equation}
Here $\mu(v_x,v_y)$ denotes the fundamental density which reads \cite{watabe:grover}
\begin{equation}
\mu(v_x,v_y) = \frac{2}{\pi^2(1-v_x+v_y)(1+v_x-v_y)(1-v_x-v_y)(1+v_x+v_y)} \mathbf{1}_{\cal E},
\end{equation}
where $\mathbf{1}_{\cal E}$ denotes the indicator function of the elliptic disc
$$
{\cal E} = \left\{(v_x,v_y)\left|\frac{v_x^2}{p} + \frac{v_y^2}{1-p}\leq 1\right.\right\}.
$$
The function $\mathbf{1}_{\cal E}$ equals 1 if the point $(v_x,v_y)$ belongs to ${\cal E}$ and zero otherwise. The symbol ${\cal M}(v_x,v_y)$ denotes the weight function which is a second order polynomial in $v_x$ and $v_y$
\begin{equation}
\label{weight1}
{\cal M}(v_x,v_y) = {\cal M}_1 + {\cal M}_2 v_x + {\cal M}_3 v_y + {\cal M}_4 v_x^2 + {\cal M}_5 v_y^2 + {\cal M}_6 v_xv_y,
\end{equation}
with coefficients ${\cal M}_j$ determined by the coin parameter $p$ and the initial coin state. Its explicit form in the standard basis is given in \cite{watabe:grover}. Finally, $\delta_0$ denotes the Dirac delta function and $\Delta$ corresponds to the localization probability around the origin. The second term in (\ref{lim:dens}) ensures that the limit density is properly normalized
$$
\int\limits_{\cal E} \nu(v_x,v_y) dv_x dv_y = 1.
$$

\begin{figure}[hbt]
\centering
\includegraphics[width=80mm,keepaspectratio,clip]{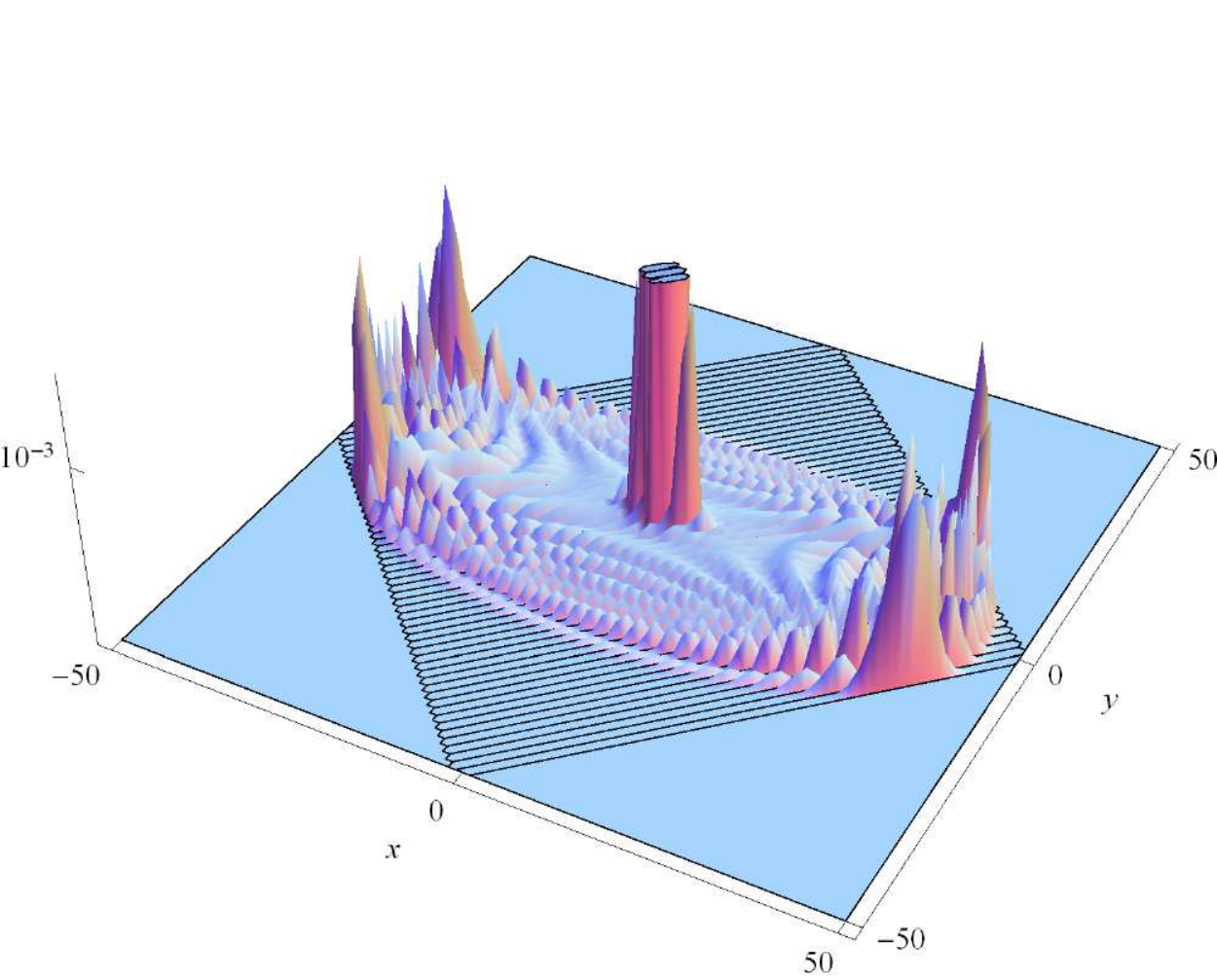}\hfill
\includegraphics[width=80mm,keepaspectratio,clip]{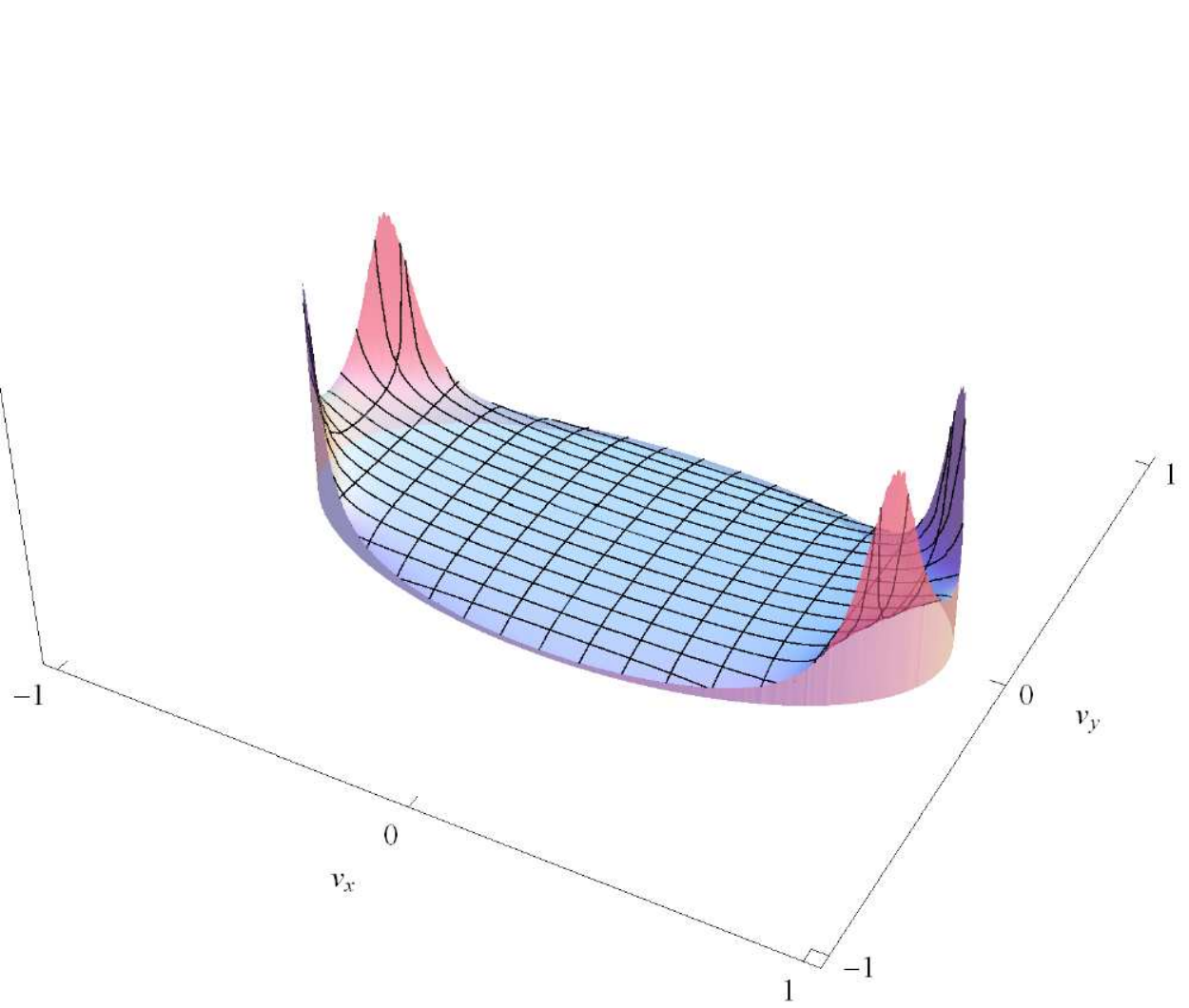}
\caption{2D quantum walk with the initial coin state $1/2(|R\rangle + |L\rangle +|U\rangle + |D\rangle)$. The coin parameter was chosen as $p=0.8$. On the left we display the probability distribution after 50 steps. The right plot shows the limit density (\ref{lim:dens}). Notice the four peaks in the probability distribution located at positions given by (\ref{peak:dist}) which correspond to the divergencies of the limit density (\ref{peak:dens}). The central peak in the left figure corresponds to the trapping probability which is not discussed in the present paper.}
\label{fig0}
\end{figure}

As we illustrate in Fig.~\ref{fig0}, generic probability distribution $w(x,y,t)$ resulting from the studied 2D quantum walk has five characteristic peaks. Four of them are propagating and after $t$ steps of the quantum walk they are located at positions
\begin{equation}
\label{peak:dist}
x=\pm p t,\quad y = \pm(1-p) t.
\end{equation}
The propagating peaks correspond to the divergencies of the limit density (\ref{lim:dens}) at points
\begin{equation}
\label{peak:dens}
v_x=\pm p,\quad v_y = \pm(1-p).
\end{equation}
These points lie at the boundary $\partial{\cal E}$ of the elliptic disc. In addition, the probability distribution $w(x,y,t)$ contains a stationary peak located at the origin. On the level of the limit density (\ref{lim:dens}) the stationary peak is described by the Dirac delta function. The peak does not vanish in the asymptotic limit $t\rightarrow +\infty$. Hence, this feature is usually called trapping (or localization), since the particle has a non-zero probability to remain close to the origin even in the limit of large number of steps. The trapping effect arises from the fact that the evolution operator of the studied 2D quantum walk has, apart from the continuous spectrum, two eigenvalues $\pm 1$ with infinite degeneracy \cite{watabe:grover}. The exact form of the trapping probability is not know, however, it decays rapidly (exponentially) with the distance from the origin. However, we will not analyze this feature in the present paper, since we focus on the properties of the limit density (\ref{lim:dens}).

In the following we consider various initial conditions resulting in non-generic probability distributions. We show that the weight function (\ref{weight1}) can be altered such that it vanishes on the boundary ellipse $\partial{\cal E}$ or on some line in the $v_x$, $v_y$ plane. Using this result we can suppress certain peaks in the probability distribution. Before we turn to the detailed analysis of the weight function we first simplify it by turning into a more suitable basis of the coin space. For this purpose we consider the orthonormal basis formed by the eigenvectors of the coin operator (\ref{coin}), which can be expressed in the following form
\begin{eqnarray}
\nonumber |\sigma_+\rangle & = & \sqrt{\frac{1-p}{2}}(|R\rangle + |L\rangle) + \sqrt{\frac{p}{2}}(|U\rangle + |D\rangle),\\
\nonumber |\sigma_1\rangle & = & \sqrt{\frac{p}{2}}(|R\rangle + |L\rangle) - \sqrt{\frac{1-p}{2}}(|U\rangle + |D\rangle),\\
\nonumber |\sigma_2\rangle & = & \frac{1}{\sqrt{2}}(|R\rangle - |L\rangle),\\
|\sigma_3\rangle & = & \frac{1}{\sqrt{2}}(|D\rangle - |U\rangle).
\end{eqnarray}
The eigenvectors satisfy the relations
\begin{eqnarray}
\nonumber C|\sigma_+\rangle & = & |\sigma_+\rangle,\\
C|\sigma_j\rangle & = & -|\sigma_j\rangle,\ j=1,2,3.
\end{eqnarray}
The initial coin state is decomposed into the eigenvector basis according to
\begin{equation}
|\psi_C\rangle = g_+|\sigma_+\rangle + g_1|\sigma_1\rangle + g_2|\sigma_2\rangle + g_3|\sigma_3\rangle.
\end{equation}
Simple algebra reveals that the coefficients of the weight function in terms of the amplitudes $g_j$ are given by
\begin{eqnarray}
\label{weight2}
\nonumber {\cal M}_1 & = & |g_+|^2 + |g_1|^2,\\
\nonumber {\cal M}_2 & = & \frac{1}{\sqrt{p}}(g_1\overline{g_2} + \overline{g_1}g_2),\\
\nonumber {\cal M}_3 & = & \frac{1}{\sqrt{1-p}}(g_1\overline{g_3} + \overline{g_1}g_3),\\
\nonumber {\cal M}_4 & = & \frac{1}{p}(|g_2|^2 - |g_+|^2),\\
\nonumber {\cal M}_5 & = & \frac{1}{1-p}(|g_3|^2 - |g_+|^2),\\
{\cal M}_6 & = & \frac{1}{\sqrt{p(1-p)}}(g_2\overline{g_3} + \overline{g_2}g_3).
\end{eqnarray}
We see that the terms ${\cal M}_1$, ${\cal M}_4$ and ${\cal M}_5$ are determined by pairs of probabilities, while ${\cal M}_2$, ${\cal M}_3$ and ${\cal M}_6$ depend on the interference of a pair of amplitudes, i.e. the coherences between the $|\sigma_j\rangle$ states. The simple form of (\ref{weight2}) allows us to identify initial coin states which lead to non-generic probability distributions in a straight-forward way.

\section{Non-generic probability distributions}
\label{sec3}

\begin{figure}[hbt]
\centering
\includegraphics[width=80mm,keepaspectratio,clip]{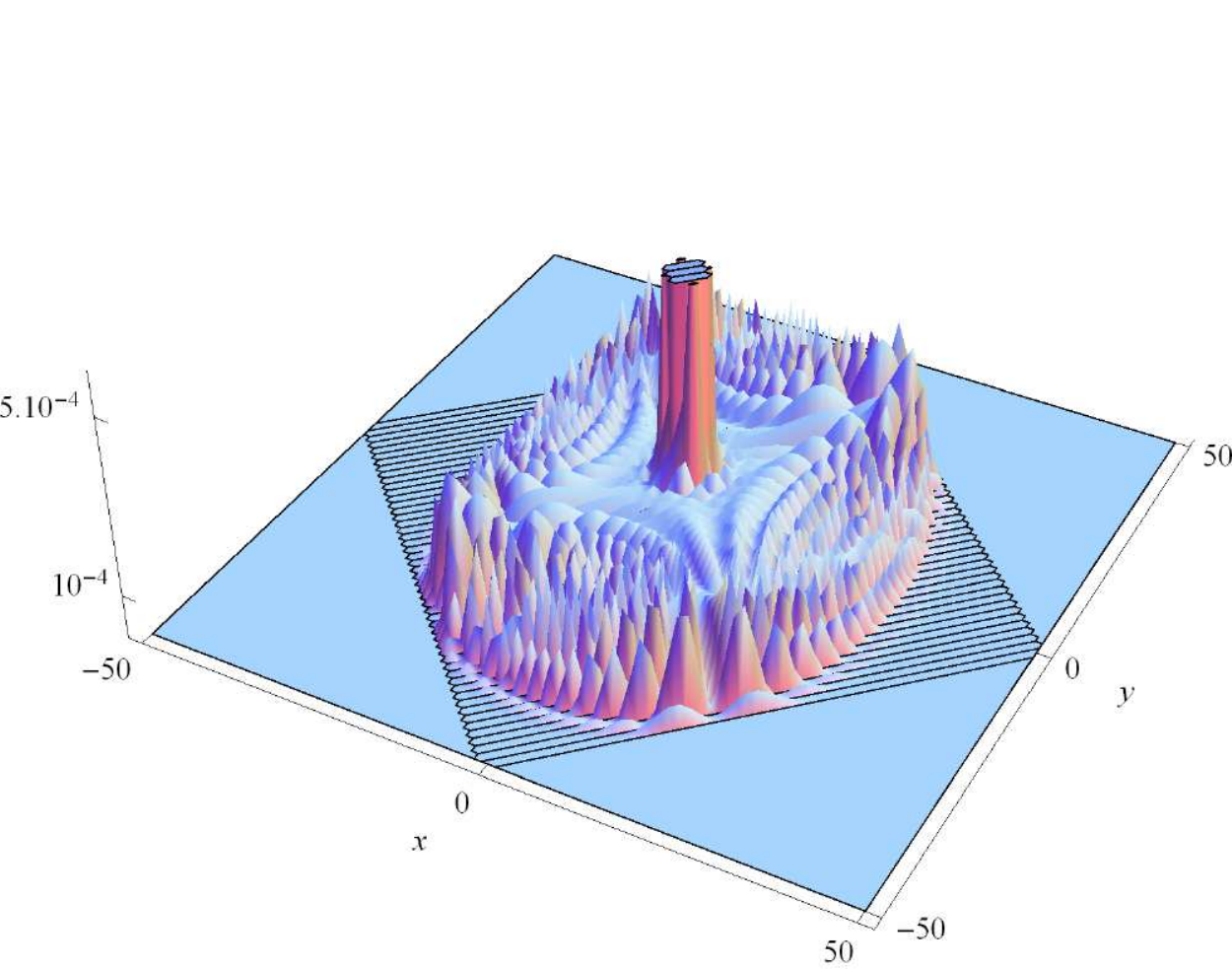}\hfill
\includegraphics[width=80mm,keepaspectratio,clip]{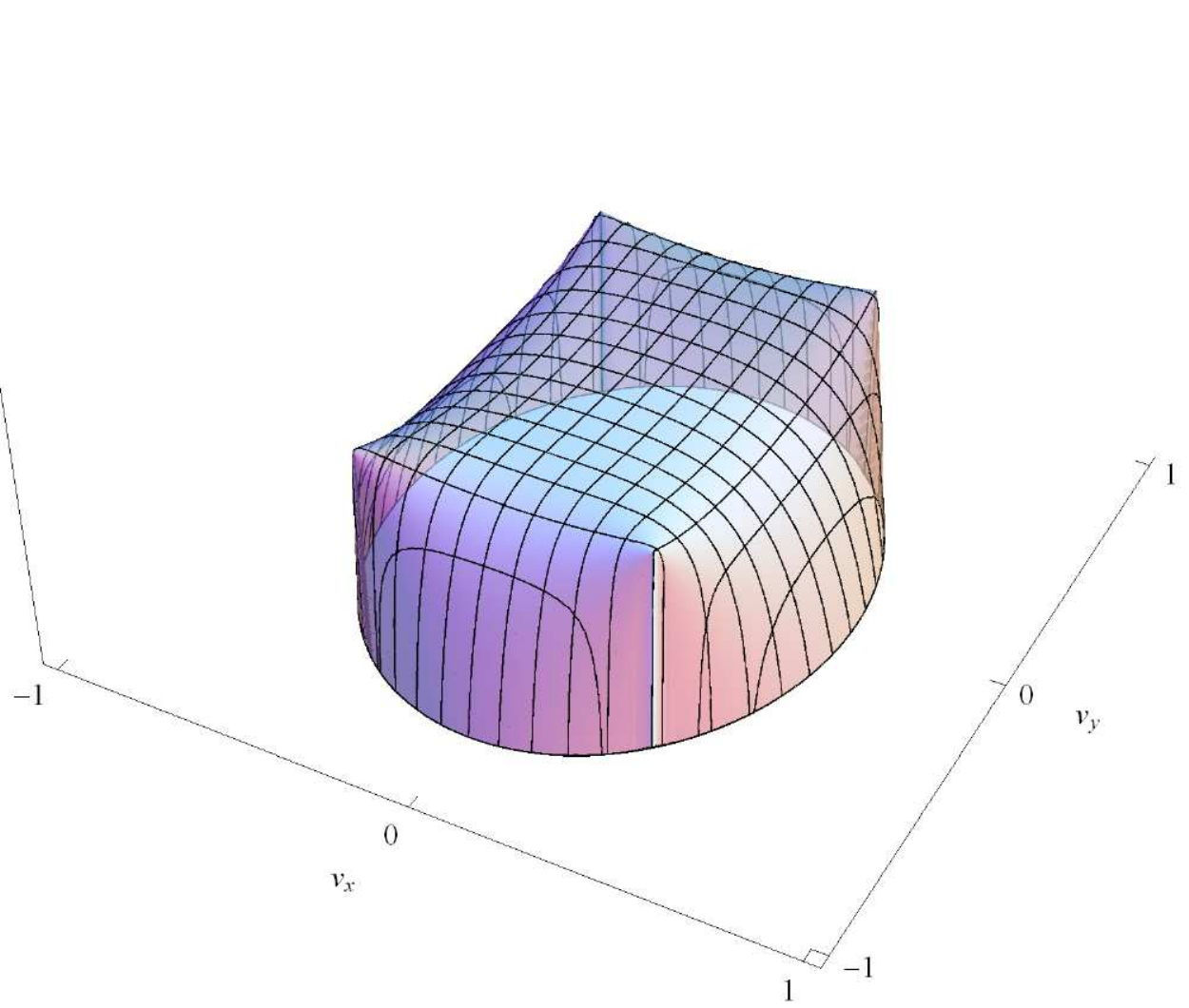}
\caption{2D quantum walk with the initial coin state $|\sigma_+\rangle$. The coin parameter was chosen as $p=0.4$. On the left we display the probability distribution after 50 steps. Notice the absence of the peaks on ellipse. Indeed, the limit density vanishes at the boundary, which we illustrate on the right. The central peak corresponds to the trapping effect.}
\label{fig1}
\end{figure}

Let us now discuss the role of the initial coin state on the shape of the probability distribution. We begin with the eigenstate $|\sigma_+\rangle$. In such a case the weight function reduces to
\begin{equation}
{\cal M}(v_x,v_y) = 1-\frac{v_x^2}{p} - \frac{v_y^2}{1-p},
\end{equation}
which vanishes on the boundary ellipse $\partial{\cal E}$. Hence, the divergencies of the limit density are suppressed and all propagating peaks will be absent in the resulting probability distribution. We illustrate this effect in Fig.~\ref{fig1}, where we choose the coin parameter $p=0.4$.

\begin{figure}[hbt]
\centering
\includegraphics[width=80mm,keepaspectratio,clip]{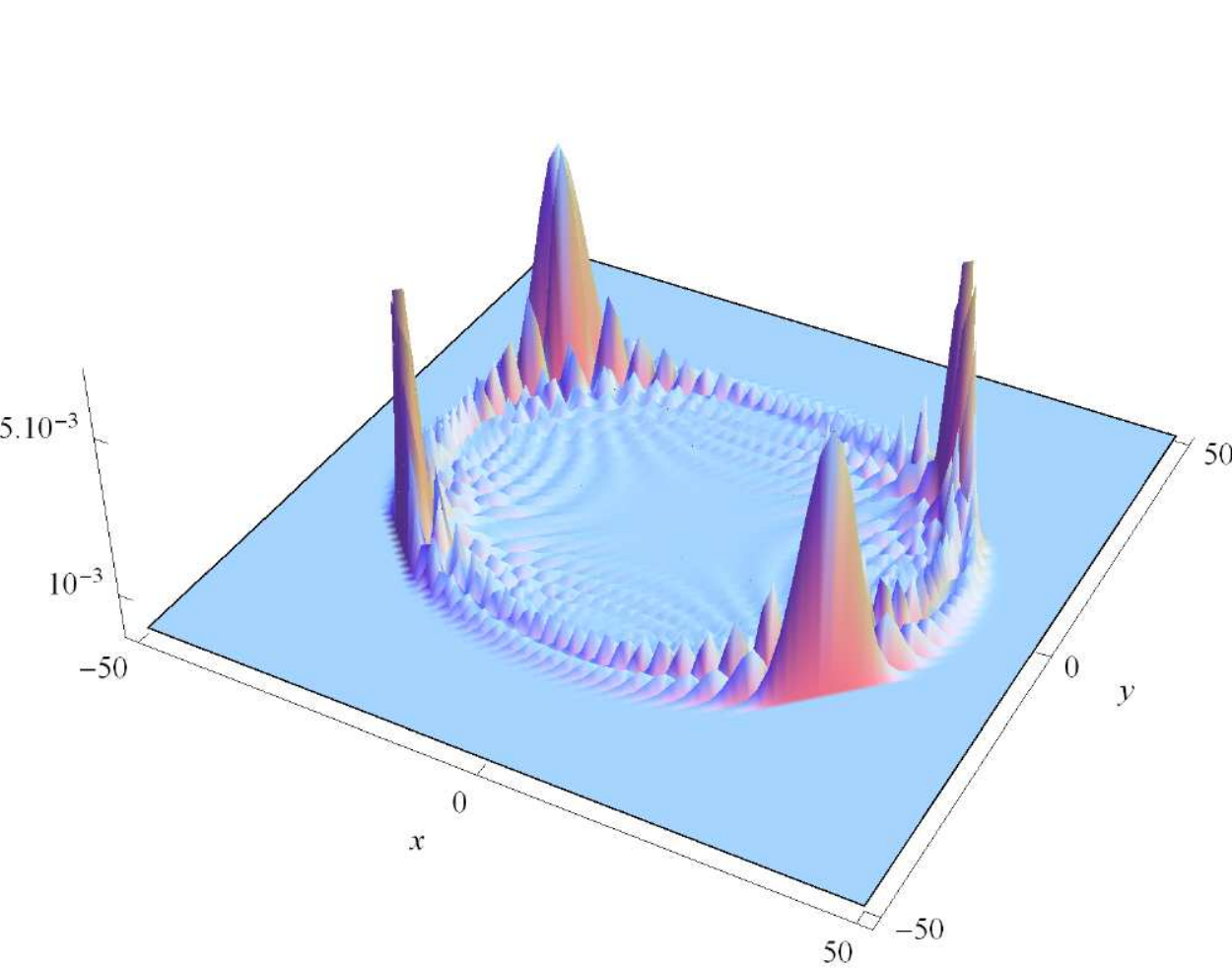}\hfill
\includegraphics[width=80mm,keepaspectratio,clip]{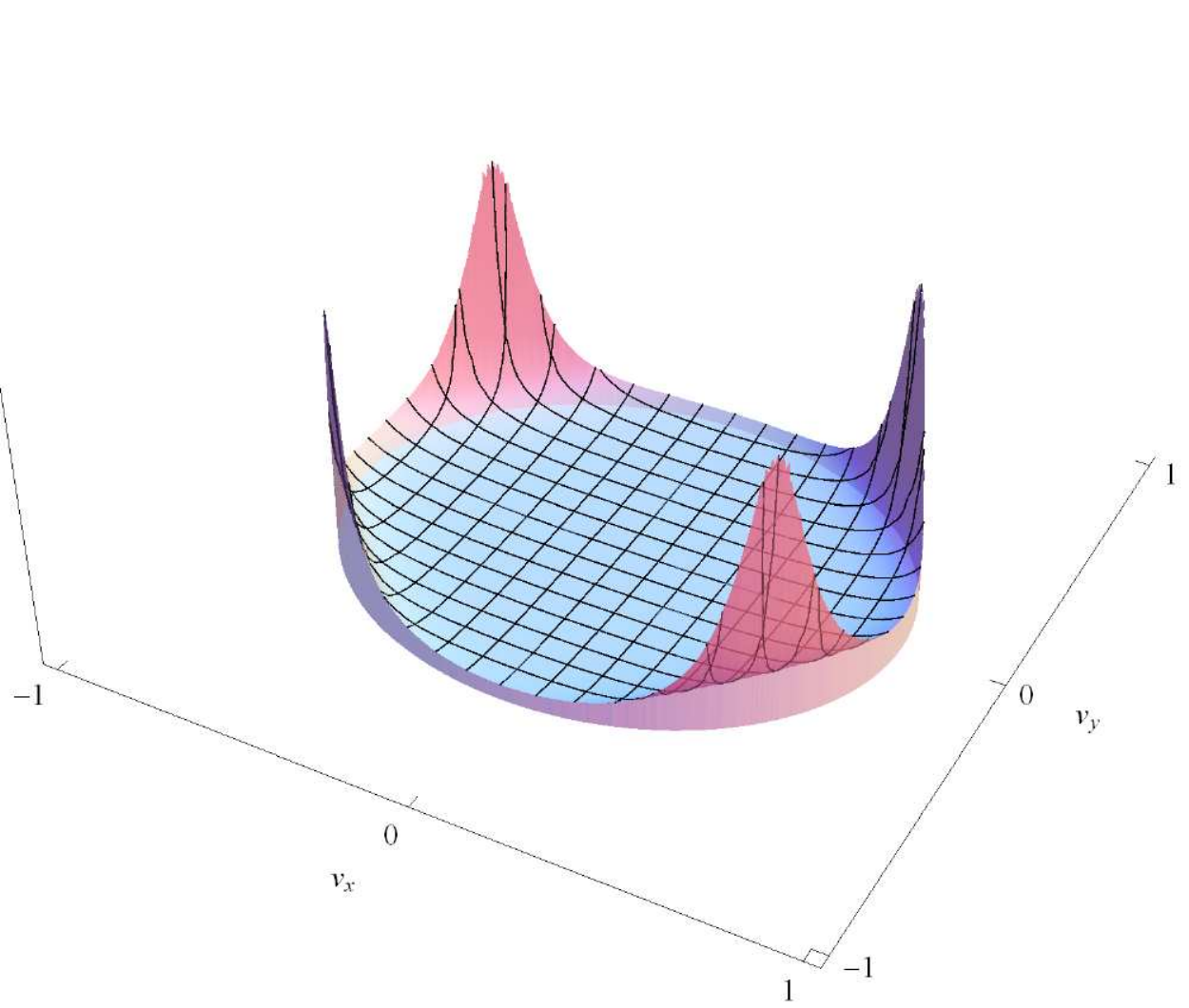}
\caption{2D quantum walk with the initial coin state $|\sigma_1\rangle$. The coin parameter was chosen as $p=0.6$. The left plot shows the probability distribution after 50 steps. Notice the absence of the central peak. Indeed, for the initial coin state $|\sigma_1\rangle$ the trapping effect vanishes. The right plot illustrates the limit density.}
\label{fig2}
\end{figure}

Next, we consider the eigenstate $|\sigma_1\rangle$. For this particular initial coin state the trapping effect vanishes, as was identified already in \cite{watabe:grover}. We illustrate this feature in Fig.~\ref{fig2} where we take the coin parameter $p=0.6$.

\begin{figure}[hbt]
\centering
\includegraphics[width=80mm,keepaspectratio,clip]{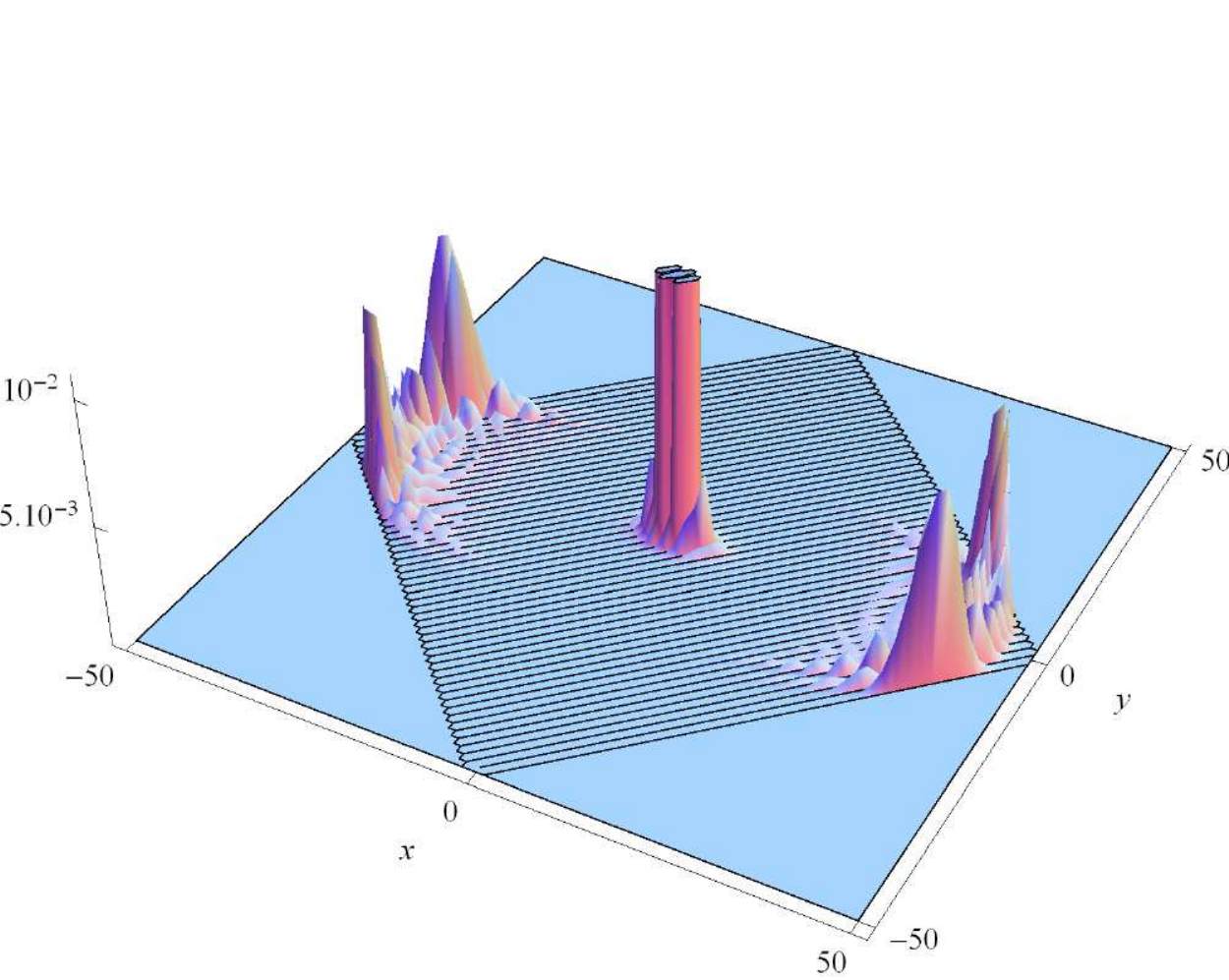}\hfill
\includegraphics[width=80mm,keepaspectratio,clip]{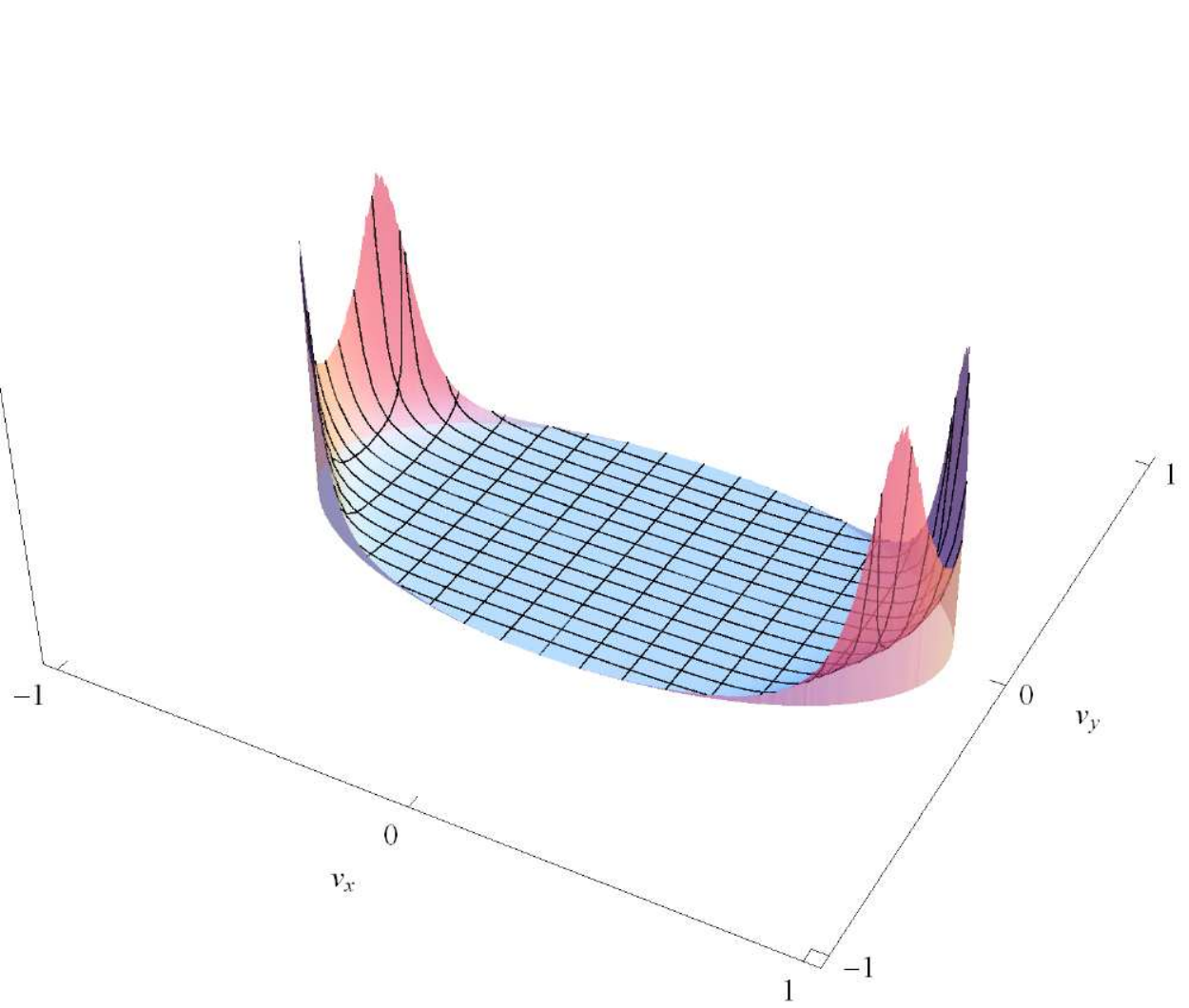}
\caption{2D quantum walk with the initial coin state $|\sigma_2\rangle$. The coin parameter was chosen as $p=0.8$. On the left we display the probability distribution after 50 steps of the quantum walk. Notice the suppression of the probability near the line $x=0$. Indeed, the limit density vanishes for $v_x=0$, as we illustrate in the right plot.}
\label{fig3}
\end{figure}

Let us now consider the eigenstate $|\sigma_2\rangle$ as the initial coin state. We find that the weight function reduces to
\begin{equation}
{\cal M}(v_x,v_y) = \frac{v_x^2}{p}.
\end{equation}
Hence, the limit density vanishes on the line $v_x=0$. This effect is illustrated in Fig.~\ref{fig3} for the coin parameter $p=0.8$.

\begin{figure}[hbt]
\centering
\includegraphics[width=0.45\textwidth,keepaspectratio,clip]{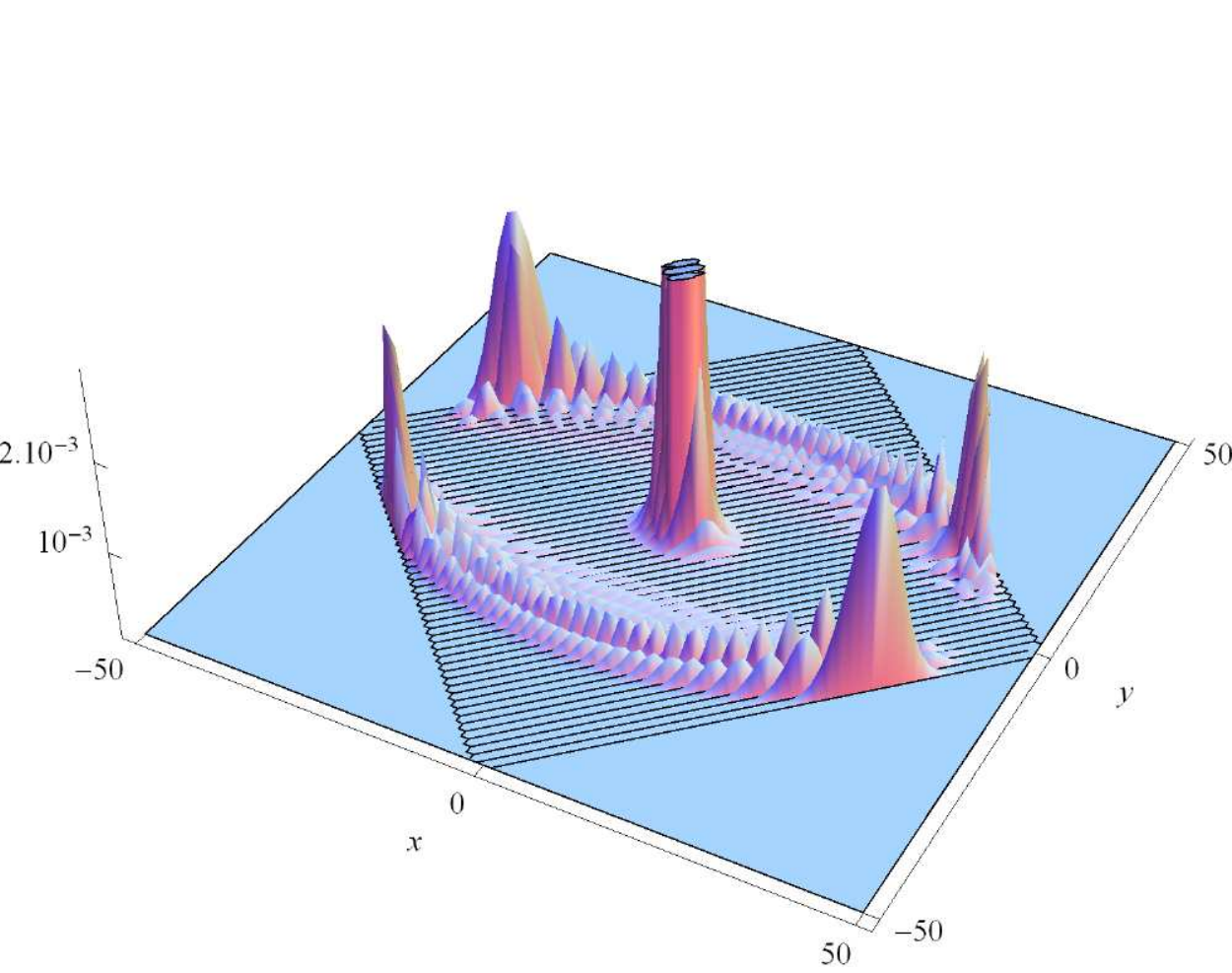}\hfill
\includegraphics[width=0.45\textwidth,keepaspectratio,clip]{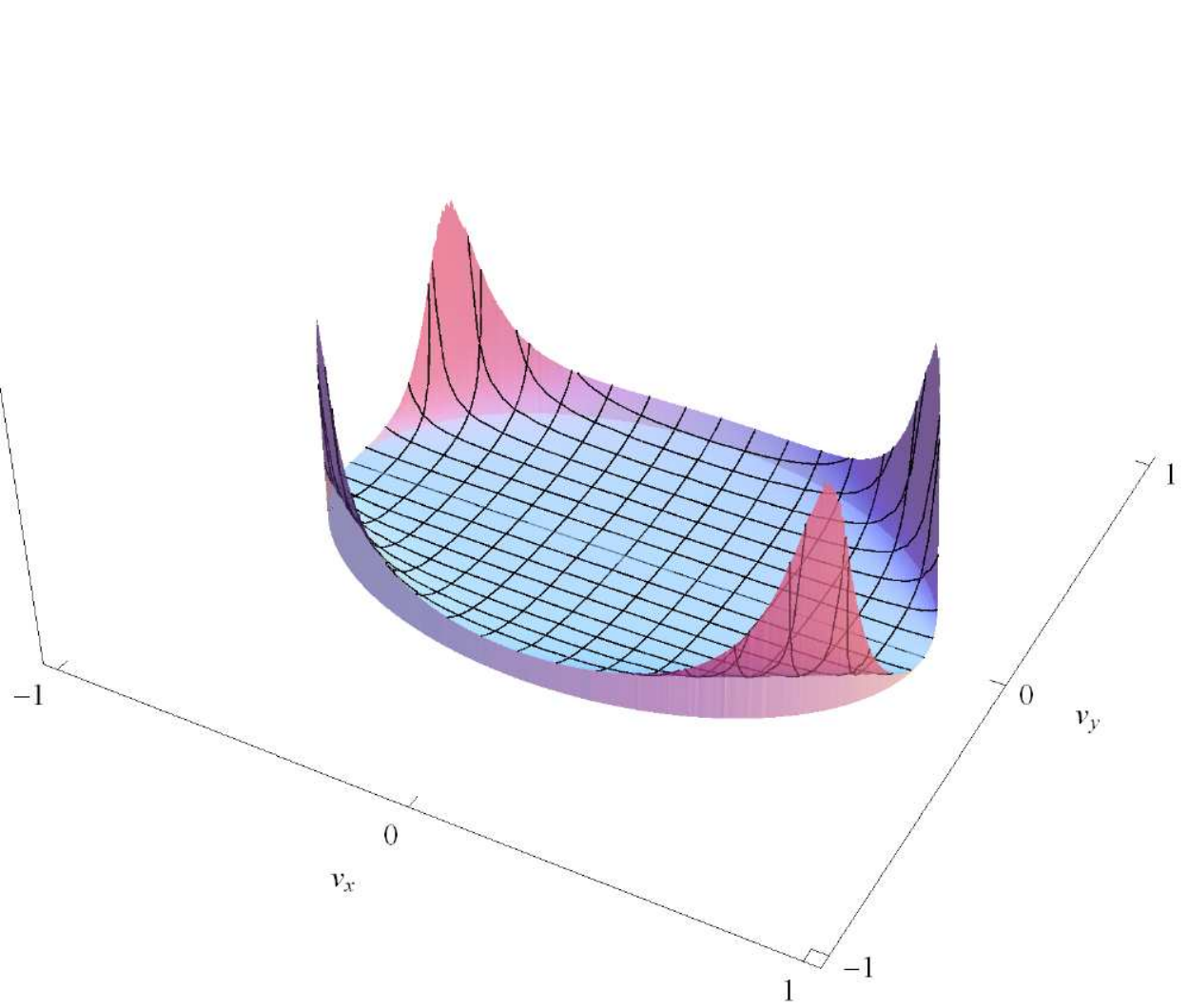}
\caption{2D quantum walk with the initial coin state $|\sigma_3\rangle$. The coin parameter was chosen as $p=0.7$. On the left we display the probability distribution after 50 steps of the quantum walk. The probability distribution is considerably suppressed along the $y=0$ line, as predicted by the limit density which is present in the right figure.}
\label{fig4}
\end{figure}

In a similar way, the choice of the initial coin state $|\psi_C\rangle = |\sigma_3\rangle$ leads to the weight function of the form
\begin{equation}
{\cal M}(v_x,v_y) = \frac{v_y^2}{1-p}.
\end{equation}
Therefore, for $|\sigma_3\rangle$ the density vanishes for $v_y=0$. This feature is depicted Fig.~\ref{fig4}.

\begin{figure}[hbt]
\centering
\includegraphics[width=0.45\textwidth,keepaspectratio,clip]{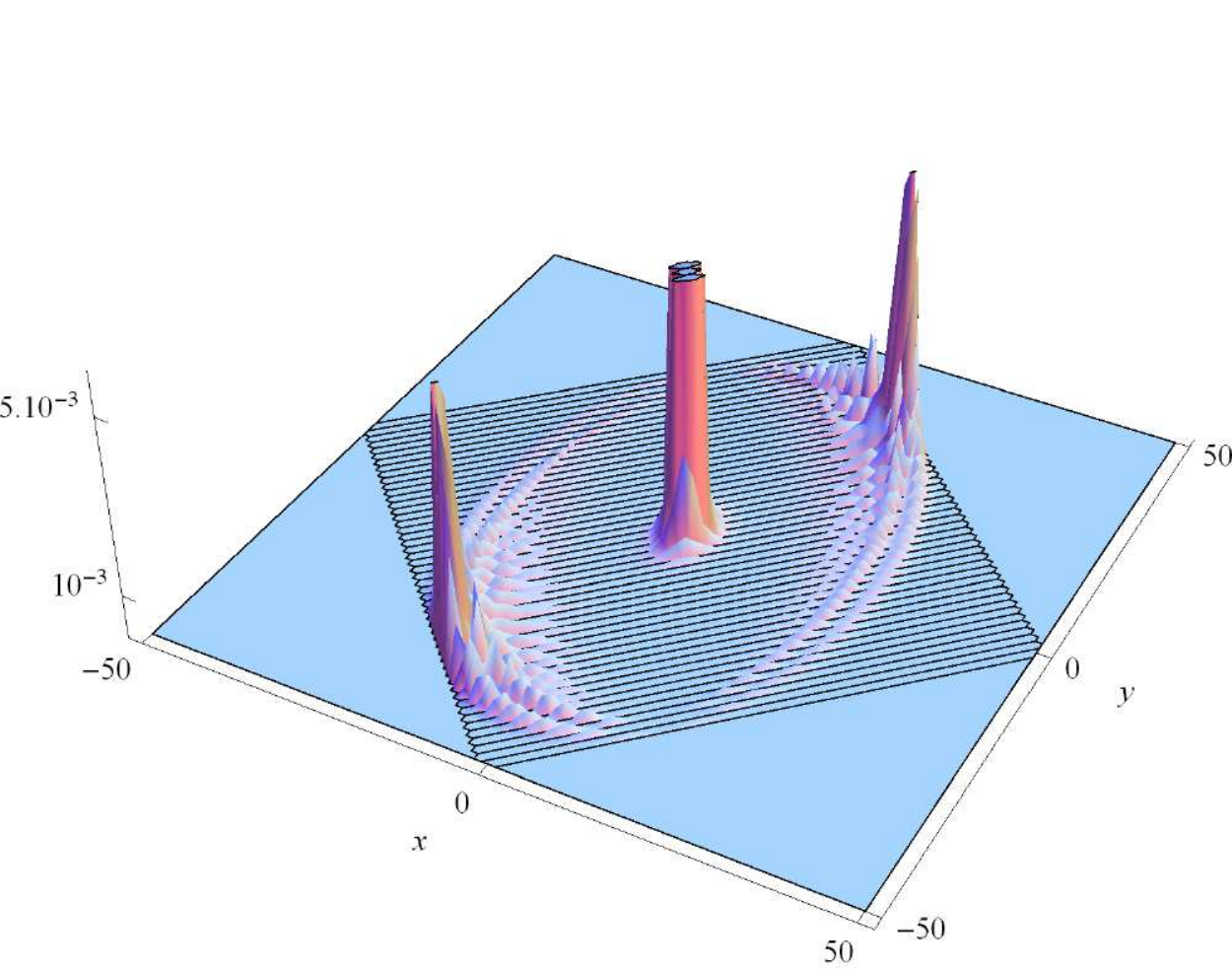}\hfill
\includegraphics[width=0.45\textwidth,keepaspectratio,clip]{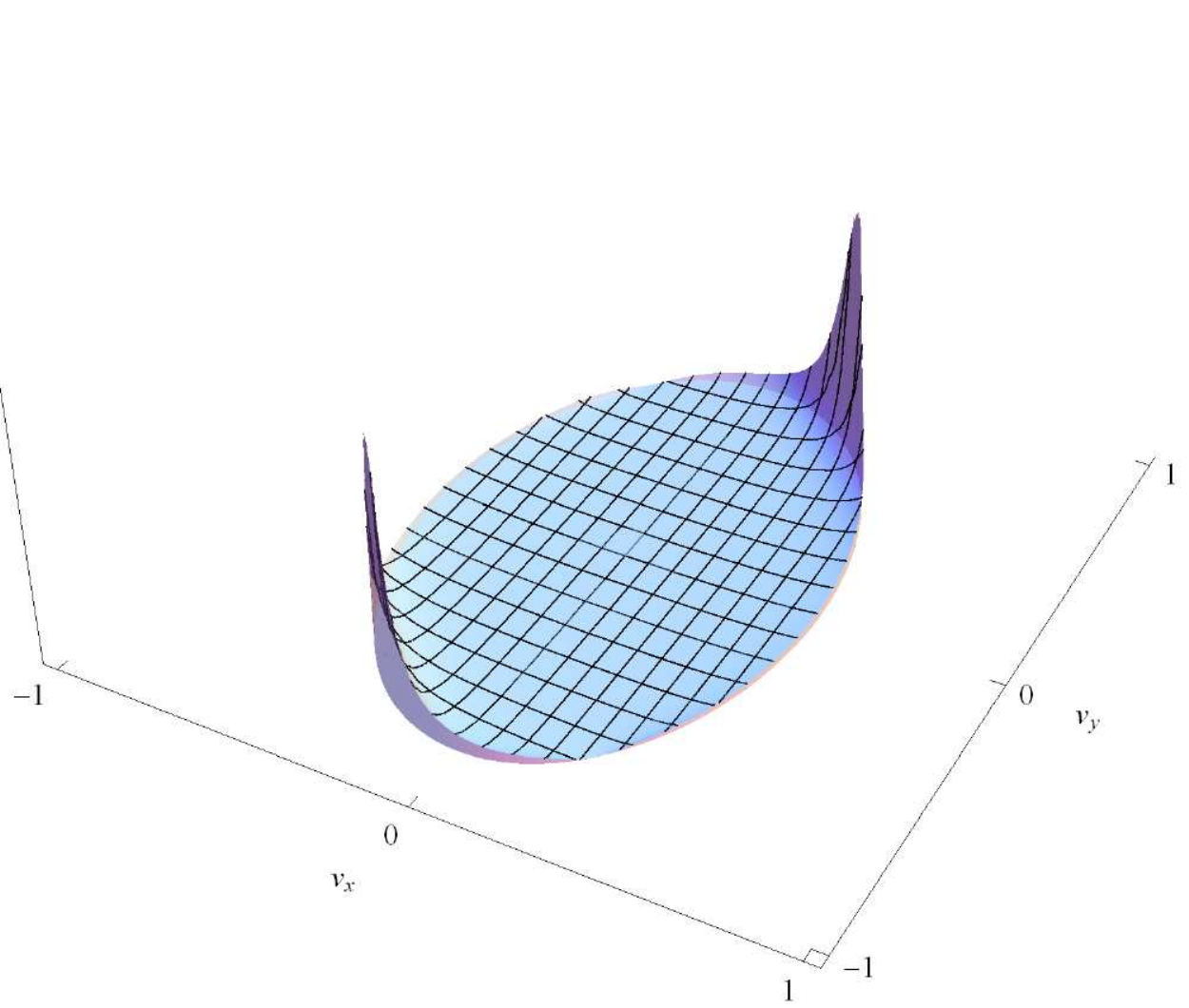}
\caption{2D quantum walk with the initial coin state given by (\ref{g2g3}). The coin parameter was chosen as $p=0.3$. On the left we display the probability distribution after 50 steps of the quantum walk. Notice that there are only two peaks on the boundary ellipse. The remaining two are suppressed since they lie on the line (\ref{line:g2g3}) where the limit density vanishes. This is illustrated in the right plot.}
\label{fig5}
\end{figure}

More generally, when we choose the initial coin state of the form
$$
|\psi_C\rangle = g_2 |\sigma_2\rangle +  g_3 |\sigma_3\rangle,
$$
the weight function reduces into
$$
{\cal M}(v_x,v_y) = \left|\frac{g_2}{\sqrt{p}}v_x + \frac{g_3}{\sqrt{1-p}}v_y \right|^2.
$$
Hence, when both $g_{2}$ and $g_3$ are real the weight functions vanishes on the line determined by
\begin{equation}
\label{line:g2g3}
\frac{g_2}{\sqrt{p}} v_x  = - \frac{g_3}{\sqrt{1-p}} v_y.
\end{equation}
We can use this fact to suppress two peaks of the probability distribution. Indeed, choosing the initial coin state as
\begin{equation}
\label{g2g3}
|\psi_C\rangle = \sqrt{1-p}|\sigma_2\rangle + \sqrt{p}|\sigma_3\rangle,
\end{equation}
eliminates the peaks at $v_x=p, v_y=-(1-p)$ and $v_x = -p, v_y = 1-p$. Similarly, for the initial coin state
$$
|\psi_C\rangle = \sqrt{1-p}|\sigma_2\rangle - \sqrt{p}|\sigma_3\rangle,
$$
the peaks at $v_x=p, v_y=1-p$ and $v_x = -p, v_y = -(1-p)$ vanishes. For illustration of this effect we display in Fig.~\ref{fig5} the probability distribution of the 2D quantum walk with the initial coin state (\ref{g2g3}) and the coin parameter $p=0.3$.

\begin{figure}[hbt]
\centering
\includegraphics[width=0.45\textwidth,keepaspectratio,clip]{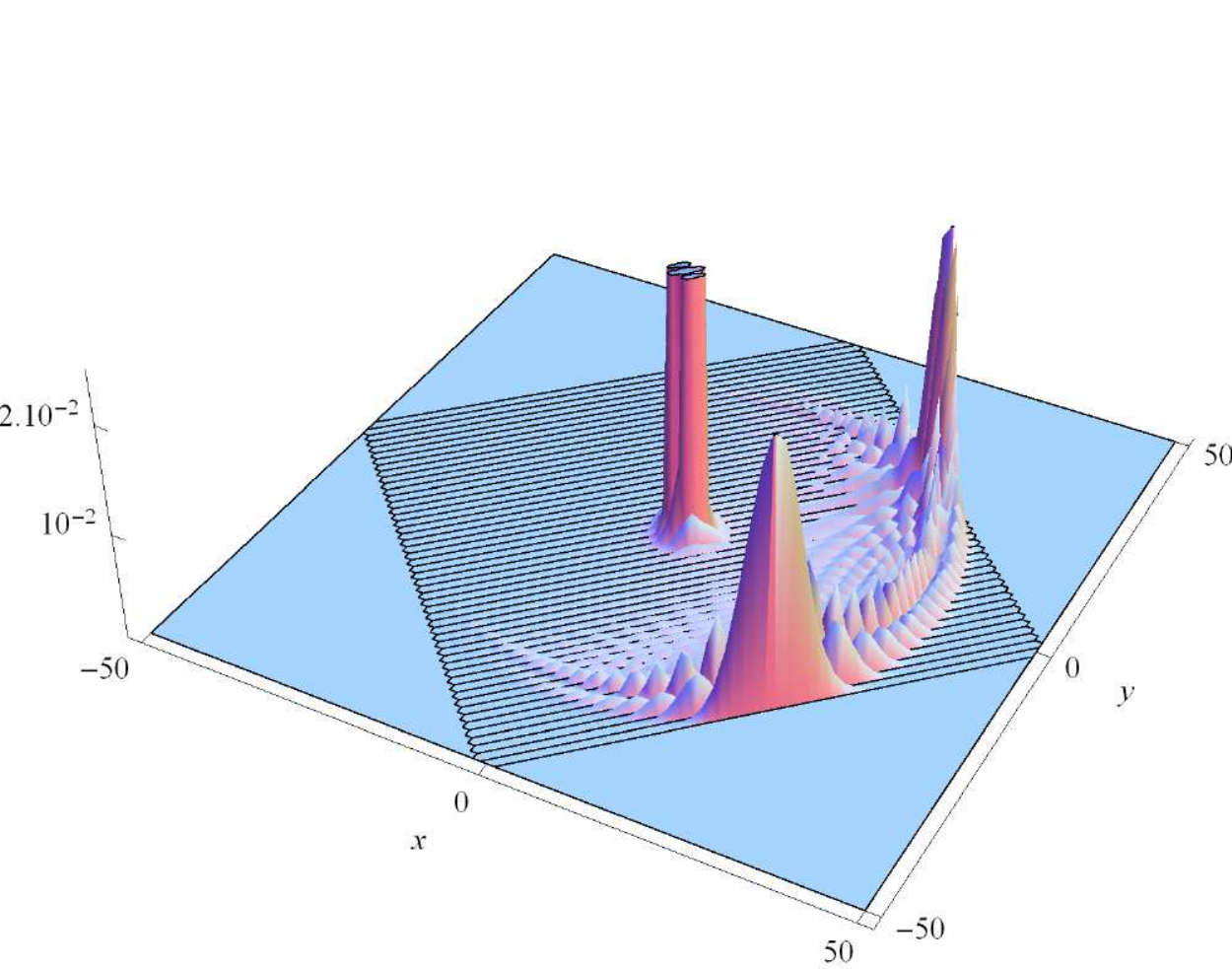}\hfill
\includegraphics[width=0.45\textwidth,keepaspectratio,clip]{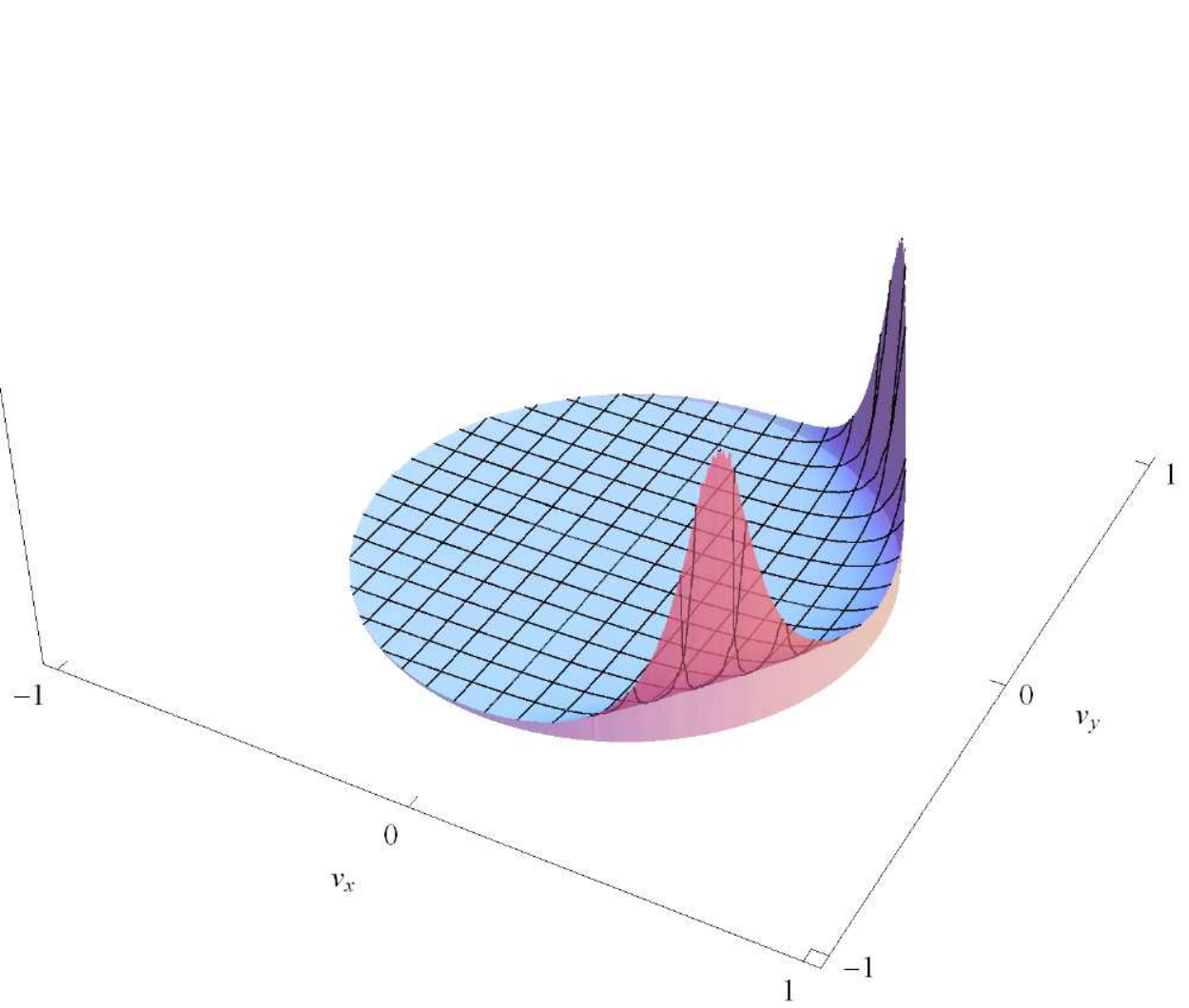}
\caption{2D quantum walk with the initial coin state given by (\ref{g1g3}). The coin parameter was chosen as $p=0.5$. On the left we display the probability distribution after 50 steps of the quantum walk. Notice that there are only two peaks on the right-hand side of the probability distribution. The remaining two are suppressed since they lie on the line $v_x=-p$ where the limit density vanishes. This is illustrated in the right plot.}
\label{fig6}
\end{figure}

Finally, we consider a situation when the weight function reduces to a polynomial only in one variable, either $v_x$ or $v_y$. We find that for $g_+ = g_3 = 0$ the weight function reduces to
$$
{\cal M}(v_x,v_y) = \left|g_1 + \frac{g_2}{\sqrt{p}}v_x\right|^2.
$$
This means that the weight function vanishes on the line
$$
v_x = -\frac{g_1}{g_2}\sqrt{p},
$$
provided that both $g_1$ and $g_2$ are real. Hence, we can eliminate the peaks on the line $v_x=\pm p$ by choosing the initial state
$$
|\psi_C\rangle = \frac{1}{\sqrt{1+p}}(\sqrt{p}|\sigma_1\rangle \mp |\sigma_2\rangle).
$$
Similarly, when we choose $g_+ = g_2 = 0$ the weight function reduces to
$$
{\cal M}(v_x,v_y) = \left|g_1 + \frac{g_3}{\sqrt{1-p}}v_y\right|^2.
$$
This means that the weight function vanishes on the line
$$
v_y = -\frac{g_1}{g_3}\sqrt{1-p},
$$
provided that both $g_1$ and $g_3$ are real. Hence, we can eliminate the peaks on the line $v_y=\pm(1-p)$ by choosing the initial state
$$
|\psi_C\rangle = \frac{1}{\sqrt{2-p}}(\sqrt{1-p}|\sigma_1\rangle \mp |\sigma_3\rangle).
$$
We illustrate this feature in Fig.~\ref{fig6} where we consider the 2D quantum walk with the initial coin state
\begin{equation}
\label{g1g3}
|\psi_C\rangle = \frac{1}{\sqrt{1+p}}(\sqrt{p}|\sigma_1\rangle + |\sigma_2\rangle),
\end{equation}
and the coin parameter $p=0.5$.

\section{Conclusions}
\label{sec4}

We have discussed in detail the role of the initial conditions on the shape of the probability distribution generated by the 2D quantum walk model analyzed in \cite{watabe:grover}. The analysis is simplified considerably by converting the results of \cite{watabe:grover} into the basis formed by the eigenvectors of the coin operator. It was found that the weight function can vanish on a certain line in the $v_x$, $v_y$ plane. Using this fact one can eliminate a pair of peaks in the probability distribution with a proper choice of the initial coin state. Moreover, the weight function can vanish on the boundary which leads to elimination of all propagating peaks.

The properties of the trapping effect were not discussed in the present contribution and remain an open question. In principle, the explicit form of the trapping probability can be obtained using similar methods as for quantum walks on a line. There it was found that the trapping probability can be highly asymmetric \cite{falkner,stef:limit}. In fact, it might be present on one half-line and vanish completely on the other. It would be interesting to see if similar features can be found in the present 2D quantum walk model.

\begin{acknowledgments}

We appreciate the financial support from RVO~14000 and from Czech Technical University in Prague under Grant No. SGS16/241/OHK4/3T/14. M\v S is
grateful for the financial support from GA\v CR under Grant No.
14-02901P. IB and IJ are grateful for the financial support from GA\v CR
under Grant No. 13-33906S.

\end{acknowledgments}

\end{document}